\documentstyle[11pt,epsfig]{article}
\newcommand{\blankline}{\vskip .3cm}
\newcommand{\f}{\begin{equation}}
\newcommand{\ff}{\end{equation}}

\begin{document}
\centerline{\LARGE Galactic disks as reaction-diffusion systems }
\blankline
\rm
\centerline{Lee Smolin${}^*$}
\blankline
\centerline{\it  Center for Gravitational Physics and Geometry}
\centerline{\it Department of Physics}
 \centerline {\it The Pennsylvania State University}
\centerline{\it University Park, PA, USA 16802}
 \vfill
\centerline{November 24, 1996}
\vfill
\centerline{ABSTRACT}

A model of a galactic disk is presented which extends the 
homogeneous  one  zone models by incorporating propagation of 
material and energy in the disk.  For reasonable values of the 
parameters the homogeneous steady state is unstable to the 
development of inhomogeneities, leading to the development of 
spatial and temporal structure.   At the linearized level a 
prediction for the length and time scales of the patterns is 
found.  These instabilities arise for the same reason that 
pattern formation is seen in non-equilibrium chemical and biological  
systems, which is that the positive and negative feedback effects 
which govern the rates of the critical processes act over different  
distance scales, as in Turing's reaction-diffusion models. This 
shows that patterns would form in the disk even in the absence of  
gravitational effects, density waves, rotation, shear and external  
perturbations. These nonlinear effects may thus explain the spiral  
structure seen in the  star forming regions of isolated flocculent  
galaxies.

\blankline
${}^*$ smolin@phys.psu.edu
\eject

\section{Introduction}

Problems of pattern formation occur on many scales in astronomy,
from the large scale distribution of galaxies to the
formation of stars and planets.  Perhaps the most studied of these
is the problem of the formation of spiral structures in galactic disks.
This has been approached by a variety of theoretical and numerical 
tools
\cite{prop,density,IBMguys,elmegreen-model,elm-trigg,elm-rev,BT}.  
However, despite the partial success of some of these models,  
there seem to be important features of the phenomena that are
so far unexplained.  These have to do mainly with 
the ubiquity,
stability and persistance of the structures seen in the star forming
regions of spiral galaxies.    The density wave models\cite{density}, 
while
succesful in some cases, have difficulty explaining why structure
should persist in isolated galaxies for times much longer than
a few rotations of the disk\cite{BT}.  On the other hand, 
models based
on the notion of propagating star formation\cite{prop}, 
while succesful in
reproducing the appearance of a range of 
galaxies\cite{IBMguys,elmegreen-model}, involve
either drastic simplifications of the astronomy or require fine
tuning of rates in order to be able to reproduce the
observed structures.  Because of these limitations, it may
not be inappropriate to consider new approaches to this problem.

Before beginning, it is important to emphasize that
the problem of spiral structure in galaxies is in fact several distinct
problems.   There is a problem of transitory structure, which is
caused by excitations of modes of the disk, most likely by encounters
with other galaxies. (This is called {\it galactic  harrasment} in
\cite{harrasment}.)  There is both observational 
and theoretical evidence
that grand design spirals are the result of such 
phenomena\cite{BT,harrasment}.  Such
transitory structure are most likely correctly seen to involve
density waves, and are outside of the phenomena to be considered
in this paper.   On the other
hand, there is a class of spiral galaxies in which the spiral structure is
primarily an aspect of the star formation process.  A typical galaxy
in this class is an  $Sc$ galaxy with flocculent spiral structure,
in the field, sufficiently far from other galaxies that the  observed 
spiral
structure must be endogenous, which is to say it must be understood 
as
a product of processes occuring in the disk.  Such galaxies typically 
show
spiral structure in blue light, 
but not in red light\cite{elm-rev,IBMguys}.  
This indicates that the
observed structure is primarily not a density wave in the disk, but is
instead a trace of the star formation process.

In such galaxies the star formation process proceeds, as far as is 
known,
at a constant rate when averaged over the 
whole disk\cite{elm-rev}. 
As a result, the
ratio of the present star forming rate to the rate averaged 
over the history
of the galaxy is one.   This constancy of the star formation rate is by
itself a significant clue, as it implies that there must be feedback 
mechanisms in the processes that govern the rate of star formation, 
so
as to keep the rate slow and 
constant\cite{feedback, parravano}.  
Given the fact that the rates are
constant over time scales much longer than those of the dynamical
processes involved ($10^{10}$ years versus at most $10^7$ years), 
there is no
other way one could explain the fact that galaxies with slow and 
steady
rates of star formation are so common.  

Other evidence concerning this will be discussed below, but the 
overall
conclusion will be that the star formation  
process can be understood to
be one component a network of self-regulated and autocatalyzed 
processes
which is the result of the
self-organization of the material in the galactic disk.  Thus, the disk 
may
be understood as a far from equilibrium statistical system in which a 
network of processes involving flows of energy and materials among 
its
several components has arisen which is governed by a set of 
feedback
loops.  Once this is seen, it is clear that the patterns produced by the
star formation process might be understood in the context of the way 
in
which spatial inhomogeneities are produced and stabilized in the
context of such non-equilibrium systems.

In recent years a body of theoretical and experimental work has 
grown up
which studies the problem of how spatial and temporal patterns
are produced in non-equilibrium 
systems 
\cite{turing,meinhardt,Stu,Harold,Prigogine,Brian,Eshel,BZ,DLA,SOC}.
The systems of interest include many
inorganic systems such as the BZ reaction\cite{BZ}, diffusion limited
aggregation\cite{DLA} and self-organized 
critical systems\cite{SOC}.  It includes as well
numerous biological systems such as colonies of bacteria\cite{Eshel},
the differentiation
of cell types\cite{Stu} 
and the formation of structure in the embryology of 
multicellular
creatures\cite{meinhardt,Stu}.   
Recently the study of such systems has produced many
successes by which patterns that can be reproduced at will in the
laboratory are explained by simple models.  These models typically
involve both partial differential equations and discrete elements 
such
as cellular automata.   As a result, one can begin to speak of the
existence of a paradigm of structure formation in far from 
equilibrium
steady state systems.  

Most of this work, and certainly the bulk of the impressive results, 
postdates
the formation of the main ideas that have governed thinking about
structure formation in astronomy.  As such it is reasonable to asses 
whether
anything has been learned there that could be of use in 
understanding 
problems of structure formation of astrophysics.  While it is 
premature
to answer generally, the purpose of this paper is to suggest that the 
answer
is yes for the particular case mentioned above, of an $Sc$ galactic
disk, far from other galaxies,  with flocculent spiral structure. 

To see why this might be the case, we may list the main elements
that characterize systems to which non-equilibrium models
of pattern formation apply.  These 
include\cite{turing}-\cite{SOC}

\begin{itemize}

\item{}The system is in a steady state, with a slow 
(relative to the relevant dynamical time scales) and 
steady flow of energy,
and perhaps matter running through it.  

\item{}The steady state is far from thermodynamic equilibrium.  
There
is a coexistence of several species or phases of matter, which 
exchange
matter and energy among themselves through closed cycles.  

\item{}The rates
at which material flows around these cycles is governed by feedback
loops that have arisin during the organization of the system to the 
steady state.  

\item{}The reaction networks of these cycles are {\it autocatalytic}.
This means that any substances that serve as catalysts or repressors
of reactions in the network are themselves produced by reactions in
the network.  

\item{}There may be spatial segretation of the different phases or
materials in the cycles.  This occurs when the inhibitory and
catalytic influences propagate over different distance scales.
At the smallest scale this means that the production of certain
substances may be subject to refractory periods, so that once 
production
has occured in a local region, it will not be repeated there for a 
certain period of time.

\end{itemize}

Given a system with these characteristics, there are models available
which describe how spatial structure is formed and stabilized.  These
come in several varieties, but the most typical are called 
reaction-diffusion systems\cite{turing,meinhardt,Stu}.  
They have been used to explain the
occurance of spatial structure in inorganic systems such as the
$BZ$ reaction\cite{BZ}\footnote{Note 
that I am not arguing that there is an
analogy between this and galactic disks because in both cases the
patterns formed are spirals.  This is a superficial resemblence, which
is due mostly to the fact that the disk is rotating differentially.  The
useful analogies are the ones I have described here.}  as well as in
organic systems such as patterns on sea shells, the stripes and spots
on the coats of mammals and the segregation of cell types in
embryology\cite{turing,meinhardt,Stu}.  

The next section begins by reviewing evidence that spiral galaxies
of the kind we are considering fit these criteria point for point.
This leads to a description of the galactic disk as an autocatalytic
network of reactions.  A first step towards a model of the
reaction network is made, which resembles a homogeneous``one zone
model".  The model is analyzed to show that the system reaches
a steady state in which  condensation of GMC's and star formation 
proceed at rates determined by a balance of positive and negative
feedback.   In section 3 and 4 this model is extended to allow
spatial variation, and it is observed that there is an alternating
pattern of negative and postive feedback as one proceeds from
the largest to the smallest scales.  Thus, the technology of
reaction diffusion models\cite{turing,meinhardt,Stu} 
is appropriate to the study of structure
formation.  The linearized analysis of this model is carried
out in section 5.  The conclusion is that for astrophysically
reasonable values of the parameters the homogeneous state
is unstable, leading to the initiation of pattern formation.  
Directions for future work are then discussed in the conclusion.

\section{Galactic disks as autocatalytic systems}

When a non-equilibrium chemical system organizes itself there 
arise cycles
of reactions by which the energy and material inputs to the 
system are
transformed into various aggregates in a steady state.  
It is, indeed, not
difficult to show that non-equilibrium steady state 
systems will generically
evolve to the point where such cycles develop.  
An important feature of
chemical reactions is also that many reactions are 
catalyzed or inhibited
by other elements.  When these catalysts or inhibitors 
are produced in the
same network of chemical reactions in which they act 
we say that the
network is autocatalytic.  Autocatalytic networks are ubiquitious
features of biological and non-equilibrium chemical 
systems.  They are
particularly important because feedback occurs naturally in 
autocatalytic 
networks.

A first step to applying the pattern formation paradigm to galactic 
disks is
to see that the interstellar medium is the site of a number of 
processes,
which form a network analogous to a network of chemical 
reactions.
Further, all of the significant processes involved in star formation are
either catalyzed or inhibited by the products of processes in the disk,
so that a galactic disk may usefully be analyzed as an example
of an autocatalytic reaction network.  

Progressing from larger to smaller scales, we may list some of 
the main reactions involved
in the star formation process, 
together with their catalysts or inhibitors.

\subsection{\bf Condensation of giant molecular clouds (GMC's)}

These cold clouds condense out of the ambient interstellar medium,
(ISM)
forming apparently scale invariant, or fractal distributions of very
cold molecular gas and dust.   

\begin{itemize}

\item{\bf Catalysts:} The main catalysts involved are dust, 
as well as carbon and oxygen.  Dust is produced mainly in
the atmospheres of cool giant stars while carbon and oxygen are
produced by fusionin stars.     The dust
shields the clouds, allowing them to cool even in the
presence of $uv$ radiation from massive stars.  The dust
grains also serve as sites for molecular binding.  The 
carbon and oxygen are
apparently necessary to cool the clouds, as radiation from rotational
modes of $CO$ is apparently the main cooling mechanism.

We may note that these substrances spread through the  $ISM$ 
over intermediate distance scale $L_{int}$, corresponding to the
scale over which the products of supernovas and massive stars
are spread through the interstellar medium by shock waves
from supernovas.  A rough scale for $L_{int}$ is
$100 $ parsecs.

\item{\bf Inhibitors:} The main inhibitor to the process is ultraviolet 
radiation from massive
stars.  This heats the ambient ISM, making condensation less 
probable.
The picture of this process proposed by Parravano and collaborators
is very useful\cite{parravano}.  They argue that there is 
a phase boundary separating
an ambient phase, in which the medium consists of warm atomic gas
and a condensed phase, in which it consists of cold molecular gas.   
The phase boundary is a curve in the $P-T$ plane, which
we may denote $T_c (P)$.  There is then
a simple feedback process which keeps the medium on the
phase boundary, at which $GMC$'s condence at a steady rate.  The
pressure of the medium is determined by the supernova rate, as the
ionized regions formed by the supernovas are the main source of 
pressurization.  Thus, the average pressure is a constant in the
steady state, and is determined by the supernova rate.  On the
other hand the temperature is due to a competition between
radiation from the disk and heating, the main source of which
is the $uv$ radiation from massive stars.  

Given this information it is clear the
system will evolve to the phase boundary.  If the massive stars heat
the medium to a temperature greater than $T_c (P)$ no more clouds
condense.  Then after a time $\tau$ on the order of $10^7$
years, a typical lifetime of a massive star, the medium will begin to
cool as a result of a decline in the $uv$ radiation as the massive 
stars that are its source 
supernova\footnote{Of course, the supernovas
heat the medium temporarily in a local region,
but over larger scales 
the heating is mainly
through the $uv$ radiation from the massive stars.  In any case if
there is no new formation of massive stars the medium must cool,
which will be after a time on the order of $10^7$ years.}.
But when the temperature
falls below $T_c(P)$ then more clouds condense, leading to the
formation of more stars and hence more $uv$ radiation.

The length scale for this inhibition process is $L_{long}$, which is
much larger than $L_{int}$ and may be as large as the radius of the
stellar disk.  The time scale is relativity short, some $10^4$ years,
which is the time it takes the radiation to spread through the disk.

We may note that the hypothesis that the condensation of the
GMC's takes place mainly on the critical curve may account for
the fact that the density distribution is scale invariant.  This
process may then be seen as an example of self-organized
criticality, by which a non-equilibrium system evolves to a steady
state in which the spatial and temporal structure is described
by power law correlations\cite{SOC}.

\end{itemize}

\subsection{The collapse of GMC cores}

The process of star formation begins when the cores of GMC's
collapse.  There may be some small spontaneous rate for core
collapse, but the collapse of cores massive enough to lead to the
formation of massive stars is apparently usually catalyzed.

\begin{itemize}

\item{\bf Catalysts:}  The main catalyst for core collapse is a shock
wave coming from either a supernova or HII regions.  
Both of these are products of massive stars.
These processes take place over the intermediate scale,
$L_{int}$.

\item{}There are also processes by which the impact of a density 
wave on a GMC may cause core collapse.  
In grand design galaxies these are thought
to be important, but they are of lessor importance for flocculent
galaxies of the type in which we are interested.

\item{\bf Inhibitors:}  The main inhibition to core collapse 
comes from two sourses.  Stellar winds from young massive
objects disrupt the molecular clouds in which they formed.  
There is also 
evaporation of the $GMC$'s due to the
$uv$ light from young massive clouds formed.  These processes
both take
place over a relativity short scale $L_{short}$ not larger than
the size of one cloud complex.  While related, this is a different
process from that by which the $uv$ light heats the ambient
medium.  We may note that these processes are effective enough
that the star forming efficiency of a given cloud is low,
of the order of a few percent.   

The effect of this process has been described in terms of a latency
time $\tau_L$.  For any given region of the disk,  the star forming
process will not generally recur for a time greater
than $\tau_L$, which is the time it takes for GMC's to begin
to condence out of the gas, after a first GMC has been evaporated
as a result of radiation produced by stars formed in it.

\end{itemize}

\subsection{Star formation}

The final stages of star formation involve the formation of a 
protostar
and acretion disk.    This process is self-limiting, according to
\cite{starforming} matter acretes onto the protostar 
until it is stopped by outflow from the young star or its accretion 
disk.  The source of this outflow may be a stellar wind from the
accretion disk, which has been heated after the ignition of nuclear
reactions.  These stellar winds also cause shock waves that may
heat and disperce the molecular clouds, further contributing to the 
inhibition of star formation.

\subsection{A model}

We may note that all of the catalysts and inhibitors involved in
the star formation process, with the sole exception of the role that
density waves play in initiating core collapse, are themselves
products of stars.  These are produced by a number of processes
by which stars return matter and energy to the interstellar medium.
We have already mentioned the most important of them 
above.  They include supernova, which 
produces shock waves, enriched material
and dust, and massive stars, which produce $uv$ light, dust 
and material
via evaporation and shock waves via the formation of ionized 
regions.  Thus, it is clear that the reactions that take place in the disk
of a spiral galaxy can be called an autocatalytic reaction network by
analogy to chemical reaction networks.

If we don't take into account spatial variation, the reaction network
may be described by a ``one-zone model" which is analogous
to the systems of equations that describe homogeneous chemical
reaction networks.
Making reasonable assumptions we arrive at a system of equations,
of the type familar from one zone models.  If we label
the densities of the components as
\begin{eqnarray}
c &\equiv& \mbox{cold gas in GMC's}  \nonumber  \\
w &\equiv& \mbox{warm ambient gas}  \nonumber  \\
s &\equiv& \mbox{massive stars}  \nonumber  \\
d &\equiv& \mbox{light stars}  \nonumber  \\
r &\equiv& \mbox{density of uv radiation} \nonumber
\end{eqnarray}
We have,
\begin{eqnarray}
\dot{c} &=& {\alpha^\prime w^2 \over r } - {\beta \over 1+ \kappa s} cs
- (\gamma + \mu ) cs 
\label{c1}
\\
\dot{s} &=&  {\beta \over 1+ \kappa s} cs -{s \over \tau }
\\
\dot{w} &=& -{\alpha^\prime w^2 \over r }+ {s \over \tau } + 
\gamma cs + \delta
\\
\dot{r}&=& \eta^\prime s - \phi^\prime wr
\\
\dot{d}&=& \mu cs 
\label{d1}
\end{eqnarray}

The model contains a number of parameters that describe the
rates of the various astrophysical processes.
$\tau$, which will turn
out to govern the characteristic time scale of the instabilities, is 
approximately the lifetime
of a typical massive star, which will be taken to be $10^7$ years.  
(More
precisely it gives the rate of flow of matter from the massive stars to 
the
warm gas, whether that takes place during the supernova or earlier 
by
evaporation.)
$\alpha^\prime$ is proportional to the 
characteristic rate for GMC's to condense from the
warm ambient gas.  As the efficiency of the formation of massive 
stars,
as well as the infall rate are small, we must choose parameters
such that in the steady state,
$\alpha^\prime w/r   > \tau^{-1}$.

Three of the parameters, $\beta , \gamma $ and $\mu$ govern the 
rates
per unit mass density at which material flows from the GMC's
to other of  the states of the
ISM by proceses which are catalyzed by the action of massive stars.  
$\beta$ and $\mu$ are  the rates for  formation of massive and light
stars, respectively,  per unit density of cold gas. $\gamma$
is the rate per unit density that the cold gas is heated by the effects 
of massive stars.   
 Finally $\delta$ is the rate at which warm gas accretes onto 
the
disk by infall from outside the galaxy.

There is an additional parameter $\kappa$
which represents the fact that massive stars cannot so easily
form near other massive stars because the gas in their neighborhood
will be heated and ionized.  We may note that there already is
a feedback effect whereas massive stars induce a heating of
cold gas to warm gas, and thus supress locally their
own formation.  However, there are also additional effects such as the 
fact that shock waves from supernova and $HII$ regions
will evaporate $GMC$'s close to the star, while they
catalyze collapse of $GMC$ cores further, after they have slowed down.
This effect is local on the scale of the model, and hence
may be represented by the parameter term $\beta /(1+ \kappa s )$ which
represents a local negative feedback for the formation of massive
stars.

The model is of course based on several simplifications.  To begin 
with the
continous spectra of stellar masses is reduced to two types.  $s$ 
measures
the density of matter in 
massive stars, by which is meant stars massive enough
to supernova, while $d$ measures the density of matter in light stars,
which are stars too light to supernova.  The fact that light stars do 
return
matter to the ISM by evaporation is one of the processes that is 
ignored 
here. Another is that there may be also a process by which
light stars condense spontaneously from the $GMC$'s.
Still another is the fact that the role of dust and carbon 
is not explicitly included, if they were this would lead to metals 
dependent corrections to the coefficients. 

In addition, the Parravano process, by which the action of the $uv$ 
radiation
is hypothesized to keep the ISM at the phase transition between 
warm
and cold gas is replaced here by a simple negative feedback, by 
which
the condensation of cold gas is suppressed by a factor of $r^{-1}$, 
which 
represents  the density of $uv$ radiation.  The reason is that 
a system with this
form of feedback evolves to a steady state at which the flows of 
material between the different
phases are constant.    A mechanism such as that proposed by
Paravanno, like any thermostat, evolves to a time dependent state 
which
fluctuates around the equilibrium point.  Thus, this kind of 
model is easier to use to study the hypothesis that instabilities
in the steady state lead to the initation of pattern formation.

A further simplification is that warm gas is assumed to fall onto
the stellar disk from outside the galaxy homogenously at a constant
rate, wheras inflow from a larger gaseous disk is possibly at least
as significant.

Even with these limitations, the system of equations
(\ref{c1}-\ref{d1})  does
give us an acceptable first model of how the
ISM arrives at a steady state at which the rates of its processes are
determined by a balance of negative and positive feedback.  Solving 
them
for the steady state at which all time derivatives  except
$\dot{d}$ vanish we find a unique solution given by
\begin{eqnarray}
c_0 &=& {1+ \kappa s_0 \over \beta \tau }   \\
s_0 &=&  {1\over 2 \kappa} 
( \sqrt{1+ {4\delta \beta \tau \kappa \over \mu}} -1) \\
w_0^3 &=& {\beta + \gamma \over \alpha^\prime \beta \tau} 
s_0^{2}
+{\delta \over \alpha^\prime } s_0
\\
r_0 &=& {\eta \over \phi^\prime  } {s_0 \over w_0 }
\end{eqnarray}
We may note the $\kappa$ independent relation
\f
s_0 =   { \delta \over \mu c_0} .
\ff
We see also that 
\f
\dot{d}=\delta
\ff
so that in the steady state forms light stars at the inflow 
rate
$\delta$.    This is of course necessary, the light stars are simply a 
sink
into which mass flows.  At a steady state the mass that flows into
the system must be 
equal to the mass that flows out of it, thus the rate of
creation of light stars must equal the rate that matter accretes or
flows onto the star forming regions of the 
disk.

\section{The hierarchy of scales of inhibition and catalysis}

System of equations such as (\ref{c1}-\ref{d1}) may 
be useful to understand why galactic
disks organize themselves into steady states  in which the 
average star formation rate is constant.  As such they may be useful
for such things as modeling the chemical evolution of the galaxy.
Of course, none of this is new, as there is a large literature on one
zone models and galactic evolution.   What we may learn from the
analogy to autocatalytic reaction networks in chemistry and biology
is that there is a natural way to understand the generation of spatial
and temporal patterns given such a reaction network.  As stated in 
the
introduction, what is needed is only that the sustances which serve
as catalysts and inhibitors spread through the system over 
different distance and time scales.  As shown in many 
examples
\cite{turing,meinhardt,Stu,Brian,Eshel,BZ,bzturbulence},
if there is a hierarchy of scales over which catalytic and inhibitory
reactions are alternatively more important, one generally
gets the formation of structure.

The main thesis of this paper is that this paradigm may be applied to
arrive at a natural understanding of the occurance of structure in
galactic disks.  To see that the preconditions are met, we may
note that in the proceeding summary of the galactic reaction
network  we have had to distinguish three distance scales.
$L_{long}$ is the scale of 
the whole stellar disk; $L_{int}$ is on the order of the
distance between cloud complexes and $L_{short}$ is the
scale of a typical cloud.  In fact, as the distribution of clouds may 
be scale invariant, these distance scales are characterized more
exactly by the processes that give rise to them: 
$L_{long}$, the heating of
the ambient medium by $uv$ light; $L_{int}$, the distance
over which a supernova may catalyze core collapse and
$L_{short}$, the range over which a new star may evaporate the
GMC out of which it condensed.

We may indeed observe that catalytic and inhibitory reactions 
alternate
in importance.  The processes that are characterized by 
$L_{int}$
are all catalytic.  This is the scale over which shock waves from
supernovas and ionization regions induce new star formation in
neighboring GMC's, and it is also the scale over which those processes
distribute the dust and enriched material produced by stars and
supernovae.

On the other hand, both the short  and the long  distance scales
are dominated by inhibitory processes.   
$L_{long}$ characterizes the heating of the ISM by $uv$ light, which
has the effect of supressing the formation of new GMC's.  
The dominant effect on $L_{short}$ is the inhibition of 
core collapse by the evaporation of a GMC by massive stars born 
there.
It is also as a result of this that the latency period is associated
with $L_{short}$.

\section{Modeling the galactic disk as a reaction diffusion system}

The conditions for structure formation in autocatalytic
networks are clearly met.  The issue is then how to model
these processes and whether all the relevant physics has been
encorporated so that the models reproduce the systematics of
$Sc$ flocculent spirals well.

The most common way to incorporate spatial inhomogeneities in
pattern formation models is through diffusion.  This is appropriate
for biological systems, and may play a role here, as there will clearly
be some diffusion of stars, clouds and materials through the disk.
However, this may not be sufficient for our purposes, as it is clear
that effects such as shock waves or the influence of $uv$ radiation
are best treated in terms of propagation rather than diffusion.

It may however be that to a first approximation propagating star
formation may be treated as a diffusion process.  This is suggested
by the phenomenological success of two different kinds of models,
in which propagating star formation is modeled.  These are the
Gerola-Seiden-Schulman model based on a cellular automota\cite{IBMguys}
and the more sophisticated Elmegreen-Thomasson 
model\cite{elmegreen-model}.  
In each of these the shock waves responsible for propagating
star formation are not modeled directly, instead one models
either a direct effect by which star-forming regions catalyze
the initiation of star formation in neighboring regions (in the
first case) or a process by which ``young stars" diffuse
from the clouds in which they are born and then give up their 
energy to nearby GMC's (in the second).  

The success of these models suggests that a diffusion-reaction
model might be applicable to structure formation in spiral 
disks.  To investigate this we may extend the homogeneous
model described above by adding diffusion terms for the massive
stars and the radiation field.  To simplify the analysis that follows,
we will reparameterize the model, by eliminating the parameters
$\beta, \gamma, \eta$ and $\mu$ in
favor of the homogeneous steady state values 
$c_0 , s_0 , w_0$ and $r_0$.
We then normalize all quantities so that 
\f
\bar{c}(x,t) \equiv {c (x,t) \over c_0 }
\ff
 and likewise for the other quantities.

We then arrive at the system of equations,
\begin{eqnarray}
\dot{\bar{c}} &=& \alpha 
\left [ {\bar{w}^2 \over \bar{r}} - \bar{c}\bar{s} \right ] 
\\
\dot{\bar{s}} &=& D_s \nabla^2 \bar{s} 
+ {\bar{s} \over \tau } 
\left (  \bar{c}({1+ \kappa s_0 \over 1+ \kappa s} )  -1 \right )
\\
\dot{\bar{w}} &=&- {\alpha c_0 \over w_0 } 
\left [ {\bar{w}^2 \over \bar{r} } - \bar{c}\bar{s} \right ] 
+ {s_0 \over w_0 \tau } \bar{s}[1-\bar{c} ]
+ {\delta \over w_0 } [1-\bar{s}\bar{c}]
\\
\dot{\bar{r}}&=& D_r \nabla^2 \bar{r} 
+\phi [\bar{s} - \bar{r}\bar{w} ]
\\
\dot{d} &=& \delta \bar{c}\bar{s}  
\end{eqnarray}

Here we have rescaled two parameters,
\f
\alpha = \alpha^\prime {w_0^2 \over c_0 r_0 }
\ff
\f
\phi = \phi^\prime w_0
\ff
Note that both now have dimensions of inverse time.  
$\alpha$ gives the
rate of condensation of cold clouds, while $\phi$ gives the
rate of energy loss of the radiation into the warm gas.

We have also introduced two diffusion constants, $D_s$ and $D_r$
which govern the diffusion of the effects of the massive stars and 
radiation, respectively.  It is reasonable to take  
\f
D_s = {L_{int}^2 \over \tau }
\ff
The value of $D_r$ may be taken to be much 
larger, perhaps on the order of 
the radius of the galaxy times the speed of light, corresponding to the
fact that the $uv$ radiation propagates through the whole disk.

As the disk is thin, it is sufficient to consider this as a model in
two spatial dimensions.

We have thus arrived at a reaction diffusion model governing the
dynamics of the interstellar medium.  Of course, it leaves out
many aspects of the physics of the galactic disk, such as gravitational
effects that may lead to density waves and external perturbations.
We have also so far not included the fact that the material in
the galaxy is differentially rotating.  

As the combinations of these
effects is already known, under appropriate circumstances, to cause
the temporary formation of spiral patterns, a complete model must
take them into account.  At the same time, we are interested in
the regieme discussed in the introduction, in which internal non-
equilibrium
processes in the disk are expected to be the cause of the observed
patterns.  This is the physics that we hope is captured in the model
we have so far.  Thus, our first task is to see if the ISM, as
described by this model, will in fact develop structure.  After we
have done this we can advance to a more complete model in which
rotation and gravitational effects are included.

Once we have the equations, the next step is to make
a linearized analysis of the theory.  This will
allow us to discover if there are unstable modes which may
develop into spatial structure.

\section{Linearized analysis of instabilities}

To proceed we now expand to linear order in perturbations from the
steady state by writing 
\f
\bar{s}= 1+S 
\ff
and likewise for the other quantities.   We arrive at a system of 
linear
equations
\begin{eqnarray}
\dot{C} &=& \alpha 
\left [ 2W - R - C - S\right ] 
\\
\dot{S} &=& D_s \nabla^2 S +  {1 \over \tau } C -S { \nu \over \tau}
\\
\dot{W} &=&- {\alpha c_0 \over w_0 } 
\left [  2W - R - C - S  \right ] 
- {s_0 \over w_0 \tau } \bar{s}C 
- {\delta \over w_0 } [ S + C ]
\\
\dot{R}&=& D_r \nabla^2 R 
+\phi [ S - R - W ]
\\
\dot{d} &=& \delta + C + S 
\end{eqnarray}

where the strength of the local saturation effect is labeled by
\f
\nu = {\kappa s_0 \over 1+ \kappa s_0}
\ff

We may now look for solutions to the linearized equations of the 
form
\begin{eqnarray}
C &=&  {\cal C} e^{\lambda t } cos(k\cdot x)  \nonumber
\\
S &=&  {\cal S} e^{\lambda t } cos(k\cdot x ) \nonumber
\\
W &=&  {\cal W} e^{\lambda t } cos (k\cdot x) \nonumber 
\\
R &=&  {\cal R} e^{\lambda t } cos(k\cdot x) 
\label{ansatz}
\end{eqnarray}
which would describe an instability with a wavevector 
$\vec{k}$ growing exponentially with a time scale $\lambda^{-1}$.
To find out if there are instabilities we must discover if there
are such solutions for reasonable values of the parameters and
reasonable wavelengths, in which
$\lambda $ is real and positive.   

Using the ansatz (\ref{ansatz}) we have an eigenvalue problem,
${\cal M}^a_{\ \ b}  V^b = \lambda V^a$, where $V^a= (C,S,W,R) $ and
the matrix is given by
\f
{\cal M}^a_{\ \ b} = \left [
\begin{array}{llll}
-\alpha & -\alpha & -2 \alpha &  -\alpha \\
1 / \tau & - k^2 D_s - {\nu \over \tau} & 0 & 0       \\
\epsilon \alpha -  {\rho \over \tau}  
- \epsilon T^{-1} &  \epsilon \alpha - T^{-1}  &
2 \epsilon \alpha & \epsilon \alpha \\
0 & \phi^\prime & -\phi^\prime & -Dr k^2 -\phi^\prime  
\end{array}
\right ]
\ff

Here we have introduced new parameters.
\f
\epsilon = {c_0 \over w_0 }
\ff
is the ratio of cold gas to warm gas in the homogeneous
solution, and should be about unity, or a bit less, as about half the
gas in the ISM is observed to be in the GMC's.   
\f
\rho = { s_0 \over w_0}
\ff
is the ratio of mass in massive stars to mass in warm gas.  
It is small, perhaps 
about $.1$ reflecting the facts that the efficiency of the
star formation process is low and the production of massive
stars is suppressed by the power law in the initial mass funcion.
\f
{1 \over T} = {\delta \over w_0} 
\ff
is the time scale for accretion of warm gas from outside the stellar
disk.  It is much longer than $\tau$, by a factor of $10^2$ to $10^3$.
 
It is straightforward to study the behavior of the eigenvalues
of this matrix as a function of the parameters.  I describe the
behavior for a typical set of parameters, and leave a more
systematic discussion for further analysis.  Some reasonable
astrophysical values for the parameters are to take
equal average densities for warm and cold gas ($\epsilon =1$),
the rate of condensation of cold gas to be about
ten times that of the return rate from massive stars to the
warm gas ($\tau \alpha =10$), $\rho$, the ratio of matter in
massive stars and warm gas at $.1$, and for the others, $T =100\tau$,
$D_r = 10^4 D_s$ and $\nu = .5$.    There is one
very negative eigenvalue, corresponding to the fast homogenization
of the radiation field, which is a result of the large value of
$D_r$.  There are then three positive eigenvalues.  This most positive
eigenvalue governs the evolution of the dominant instability
of the disk.  In Figure 1 we graph the real part of the
largest positive eigenvalue as a
function of the wavelength $l= 1/|k|L_{int}$.  We see that the
most unstable modes are at short scales on the order of 
$L_{int}$.  We don't trust the model for smaller scales, so
what we can say from the graph is that with the
parameters as chosen the model generates
instabilities over all scales, with the fastest
growing instabilities between $L_{int}$ and
about $20 L_{int}$.  

\begin{figure} \centerline
 {\mbox{\epsfig{file=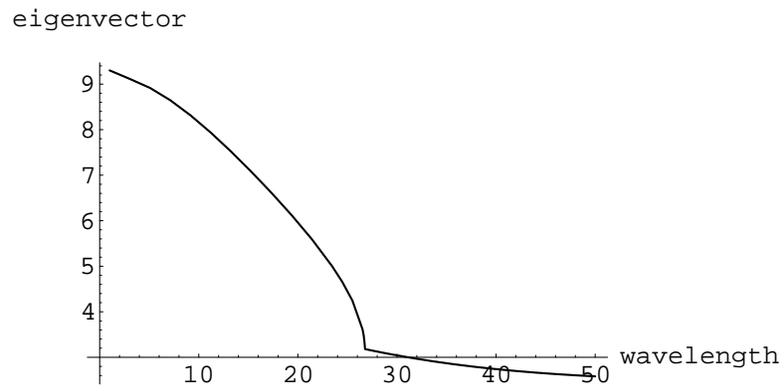}}}
\caption[Fig. 1.]{Dependence of the most unstable eigenvalue on
wavelength, in units of $L_{int}$.  The eigenvalue is in units of
$10^{-7} sec^{-1}$.}
\end{figure}
 
The time scale of the instability is about $10^6$ years.
Thus, within the limitations of the model, we may
say that the scales of these
instabilities are right to be the initiation of the formation
of spiral patterns as seen in the flocculent galaxies. 
However  to
understand the formation and evolution of these patterns beyond 
this
initial stage we need to carry out the full non-linear analysis.

\section{Conclusions}

The results of the linearized analysis support the basic theses
put forward in the earlier sections.  First, that a galactic disk
can be seen as an autocatalytic network of reactions in which
structure forms and persists because catalysis and 
inhibition (or postive and negative feedback) alternatively
dominate as one decends from larger to smaller length
scales.  This means  that patterns spontanteously
form in the medium, even before we take into account
gravitational effects, density waves, rotation or external
perturbations.
This suggests that the general framework of reaction-
diffusion models may be appropriate for an understanding
of the phenomenology of spiral galaxies.  Further, the time
and length scales of the instabilities are reasonable.  

However, these results must be considered preliminary.
Rotation
must be added to the model, after which a full
non-linear analysis must be carried out for a variety of parameters,
and compared with structure observed in galaxies.    

One flaw of the model which is visible even at the linearized level
is that it does not allow a clear separation between local 
inhibitory effects and intermediate scale catalytic effects.
Thus,  near massive stars
uv radiation and shock waves act to destroy GMC's, while further
away at the intermediate scale the shock waves catalyze further
star formation.  As pointed out by Anderas Freund \cite{af}, this
distinction is not represented in the model, which results in the
fact that there is no supression of the instability at short scales.
This  defect can be remedied by the explicit introduction of shock waves,
so that the local effect of the star on its sorroundings can be
cleanly separated from the effect of its shock waves at intermediate
scales.
Preliminary results\cite{af} show that when this is done there is a
clear separation of small scale supression from medium scale
excitation resulting in a short distance cutoff for the scale of
the instabilities.   

Even with these improvements  it may be
expected that this kind of approach could successfully describe
only  the type of galaxies
mentioned above, (flocculent 
Sb or Sc spirals, isolated from other galaxies) as
these are the cases in which spiral structure is seen only in the
star forming regions and not in the density of old stars.   Still 
success in this domain would support the theses of this paper.
While other models of these kinds of galaxies exist which do
successfully reproduce the observed 
structure\cite{IBMguys,elmegreen-model} these involve
either drastic simplification of the physics or 
fine tuning of parameters.  The model proposed here, if successful
would explain how the system of the galactic disk organizes itself
to tune its own parameters to a state characterized by a  constant
overall rate of star formation in which the star forming
regions produce slowly evolving spiral patterns.

To gain a complete understanding of the dynamics of spiral
disks, the model will have to be
extended to include gravitational and density effects as well
as rotation.  The
goal in the end will be an understanding of how effects of
gravitation and rotation couple with the processes described here
to produce the whole range of types of spiral structures.   In
particular, it might  very well be the case that turbulence in
the interstellar medium plays a role in star formation over a
range of scales\cite{elmefr}.  However this may not be as much
a competing hypothesis as complementary to the ideas discussed
here, given the fact that a transition has been observed in the 
BZ system\cite{BZ} in which perfect spirals are disordered by defects
leading to a kind of turbulent behavior reminicent of the patterns
seen in flocculent galaxies\cite{bzturbulence}.

Finally, it may be that similar methods could lead to an
understanding of other astrophysical problems involving
pattern formation such as non-linear processes in galaxy formation.

\section*{ACKNOWLEDGEMENTS}

I am grateful especially to Stuart Kauffman for telling me about
reaction diffusion models and for much insight and  discussion 
and to Anna Jangren for information and discussions during the course 
of this work.   Conversations with Andreas Freund have also
been extremely helpful.   I learned a 
great deal of astronomy from  conversations with Jane Charlton, 
Bruce Elmegreen, Greg Stevens, Pete Hut, Peter Meszaros, 
Martin Rees, 
Peter  Saulson and Larry Schulman.  Likewise my understanding of 
reaction diffusion equations was greatly helped by talking to Brian 
Goodwin, Marcelo Magnasco and Hans Meinhardt.  
Criticisms of an earlier draft of this paper from 
John Baker, Bruce Elmegreen
Andreas Freund, Anna Jangren and Peter
Meszaros were also most useful.
This work was 
begun during a visit to the Santa Fe Institute 
and continued during visits to Rockefeller University and SISSA
in Trieste.  The work was supported by the National Science
Foundation under grants PHY93-96246 and PHY95-14240.  
Finally, I thank the guys for help with
Mathematica.

\end{document}